\documentclass[preprint,aps,draftams,math,amssymb,superscriptaddress,unsortedaddress]{revtex4}
\usepackage{graphicx}
\usepackage{dcolumn}
\usepackage{bm}
\usepackage[usenames,dvips]{color}
\usepackage{gensymb}
\usepackage{amsmath}
\newcommand{\comment}[1]{}
\begin{document}

\preprint{APS/123-QED}

\title{A unified cluster expansion method \\applied to the configurational thermodynamics of cubic TiAlN}

\author{B. Alling}
\email{bjoal@ifm.liu.se}
\affiliation{Department of Physics, Chemistry and Biology (IFM), \\
Link\"oping University, SE-581 83 Link\"oping, Sweden.}
\author{A.~V. Ruban}
\affiliation{Department of Materials Science and Engineering, \\
Royal Institute of Technology (KTH), SE-100 44 Stockholm, Sweden.}
\author{A. Karimi}
\affiliation{Institute of Condensed Matter Physics (IPMC),\\
\'Ecole Polytechnique F\'ed\'erale de Lausanne (EPFL), CH-1015 Lausanne,
Switzerland }
\author{L. Hultman}
\affiliation{Department of Physics, Chemistry and Biology (IFM), \\
Link\"oping University, SE-581 83 Link\"oping, Sweden.}
\author{I.~A. Abrikosov}
\affiliation{Department of Physics, Chemistry and Biology (IFM), \\
Link\"oping University, SE-581 83 Link\"oping, Sweden.}
\date{\today}

\begin{abstract}
We study the thermodynamics of cubic Ti$_{1-x}$Al$_x$N using a unified cluster
expansion approach for the alloy problem. The purely
configurational part of the alloy Hamiltonian is expanded in terms
of concentration and volume dependent effective cluster interactions. 
By separate expansions of the chemical fixed-lattice, and local lattice relaxation terms of the ordering energies,
we demonstrate how the screened generalized perturbation method can be fruitfully combined with a concentration dependent
Connolly-Williams cluster expansion method. 
Utilising the obtained Hamiltonian in Monte Carlo simulations we access
the free energy of Ti$_{1-x}$Al$_x$N alloys and construct the isostructural
phase diagram. The results show surprising similarities with the previously
obtained mean-field results: The metastable c-TiAlN is subject
to coherent spinodal decomposition over a larger part of the
concentration range, e.g. from $x\geq 0.33$ at 2000~K.

\end{abstract}

\maketitle

\section{Introduction}
Metastable cubic Ti$_{1-x}$Al$_x$N solid solutions~\cite{Beensh1981,Knotek1986,Munz1986} are one of the most widely
used classes of hard protective coating materials~\cite{Mayrhofer2006Rev}. The addition of Al to TiN
coatings has been shown to result in better oxidation resistance~\cite{Knotek1986}
and retained or even increased hardness at temperatures reached in for cutting
tool applications~\cite{Mayrhofer2003}. Al-rich cubic TiAlN films demonstrates
particularly good properties. The age-hardening mechanism has been shown to be due
to a coherent isostructural spinodal decomposition of the as-deposited cubic
Ti$_{1-x}$Al$_x$N solid solutions into TiN (or Ti-rich TiAlN) and cubic AlN
domains~\cite{Mayrhofer2003, Santana2004}. The existence of an energetic
de-mixing driving force has been proven theoretically by means of first-principles
calculations~\cite{Mayrhofer2006, Alling2007} and explained to be primarily due
to an electronic structure mismatch effect particularly strong for the Al-rich
compositions~\cite{Alling2007}.

The isostructural phase diagram, derived theoretically within the
mean-field approximation~\cite{Alling2007}, shows an almost complete miscibility
gap between TiN and c-AlN at temperatures of relevance for cutting tool applications, around $1300~K$. According to
these estimates, the composition region displaying spinodal decomposition extends
from $x=0.25$ to $x=0.99$ at this temperature~\cite{Alling2007}, but could be
further extended when subjected to cutting-induced pressure~\cite{Alling2009APL,
Holec2010}. However, the mean field approximation neglects the effect
of short range order or clustering and is known to overestimate the critical
temperatures~\cite{Ducastelle1991}. The qualitative importance of local clustering in
TiAlN for the isostructural decompositions as well as for the maximum amount
of Al possible to solve in the cubic phase before appearance of wurtzite AlN
during synthesis has been noted in a number of papers~\cite{Mayrhofer2006,Mayrhofer2006JAP,Alling2007}.

Here, we investigate the clustering thermodynamics of cubic TiAlN using
an accurate statistical mechanics analysis beyond the mean field
approximation. In doing so we use concentration and volume dependent effective
cluster interactions, which are devised by a here proposed combination of two complementary methods:
the generalized perturbation method (GPM) ~\cite{Ducastelle1976} and the
structure inverse or Connolly-Williams cluster expansion (CE) method.~\cite{Connolly1983} In this
way we are able to provide accurate mapping of the complex interatomic interactions
onto an Ising-type Hamiltonian, which can then be used in Monte Carlo
calculations of the configurational part of the free energy and its contribution
to the isostructural phase diagram of the system. 

\section{Alloy Hamiltonian}

In a previous investigation of the c-Ti$_{1-x}$Al$_{x}$N system,
the electronic structure and as consequence, mixing energy of these alloys
was found to exhibit complicated behavior, which could not be captured by a simple regular
solution model. Such a behavior of the system, and the changes of its thermodynamic 
properties with concentration, is related to a gradual electronic structure transition due to bond cutting of 
the next-nearest neighbor (metal sublattice nearest neighbor) Ti-Ti bonds of $t_{2g}$ character, eventually leading to isolated
and localised states in a semiconducting AlN-rich matrix~\cite{Alling2007}.
This is in fact exactly
the case where the concentration independent (grand canonical) cluster expansion
brakes down: the nature of Ti-Ti interaction in the metallic state and in the semiconductor state 
are different.


Therefore, the concentration dependent expansion of the configurational energy is
the only reasonable choice for the alloy Hamiltonian in TiAlN. In general such
a Hamiltonian is defined as

\begin{eqnarray}\label{eq:H_conf}
 \mathcal{H}_{conf} &=&
 \frac{1}{2}\sum_{p} V^{(2)}_p(x) \sum_{i,j\in p} \delta c_i \delta c_j+
{}\nonumber \\
&+& \frac{1}{3}\sum_{t} V^{(3)}_t(x) 
  \sum_{i,j,k\in t} \delta c_i \delta c_j \delta c_k+ {}\nonumber \\   
&+& \frac{1}{4}\sum_{q} V^{(4)}_q(x)
  \sum_{i,j,k,l\in q}\delta c_i \delta c_j \delta c_k \delta c_l+{}\cdots
\end{eqnarray}
\noindent where $V^{(n)}_\alpha(x)$ are the $n$-site
effective cluster interactions for specific cluster $\alpha$ and the
concentration of Al $x$; $\delta c_i = c_i - x$ are the concentration
fluctuation variables defined in terms of the occupation numbers $c_i$,
which takes on values 1 or 0, if Al or Ti atom occupies site $i$,
respectively.

The Hamiltonian in Eq. (\ref{eq:H_conf}) takes care only of the configurational part,
so the total alloy Hamiltonian can be obtained if either the total energy
of a random alloy or its mixing enthalpy (in this case at zero pressure), $\Delta H_{mix}(x) = E_{tot}(x) -
x E_{tot}(x=0) - (1-x) E_{tot}(x=1)$, are added to
the configurational Hamiltonian. In this work we use the latter for 
convenience: 

\begin{equation} \label{eq:H_alloy}
\mathcal{H}_{all} = \Delta H_{mix}(x) + \mathcal{H}_{conf} .
\end{equation}
Eq.~(\ref{eq:H_alloy}) follows from the fact that the configurational energy of a
random alloy in Eq. (\ref{eq:H_conf}) is exactly zero, since in this case
$<\delta c_i \delta c_j \cdots \delta c_k > = 
<\delta c_i> < \delta c_j> \cdots <\delta c_k > = 0$. Let us note, that
in the case of the concentration independent cluster expansion, the
mixing enthalpy is re-expanded in terms of the corresponding effective
cluster interactions.

The next important point is the separation of the effective cluster
interactions in the Hamiltonian of Eq.~({\ref{eq:H_conf}) into chemical part,
defined as the situation with all atoms sitting on fixed ideal lattice points in the absence of any local relaxations, and the 
additional relaxation
part, which takes care of the rest of the energy, associated with local
lattice relaxations specific for a given atomic configuration:

\begin{equation}\label{eq:V}
V^{(n)}_{\alpha} = V^{(n)-fix}_{\alpha} + V^{(n)-rel}_{\alpha} ,
\end{equation}
where $V^{(n)-fix}_{\alpha}$ and $V^{(n)-rel}_{\alpha}$ are the chemical fixed-lattice
and relaxation part of $n$-site interaction $V^{(n)}_{\alpha}$, respectively.

With this separation one can gain an important advantage if there
is an efficient way to get the chemical part of the interactions. This part may exhibit non-trivial behavior and nonlinear concentration
dependence, induced for instance by the electronic structure effect as demonstrated in Ref.~\cite{Alling2007}. 
On the other hand, at least in the system studied here, the local lattice relaxations, although important, show a very
smooth behavior as a function of concentration. In fact, in TiAlN the latter
originates to a large extent from the relaxation of nitrogen atoms and can
be well explained by the independent sublattice model, given by a simple
one-parameter equation~\cite{Alling2007}. In this work, however, we aim for a more detailed description also of the relaxation part of the problem.

The effect of thermal expansion as well as other vibrational effects and electronic excitations are neglected in this work.
We do not believe that thermal expansion has a large impact on the here calculated properties in the temperature regions that have any practical relevance.
This is so since in a test calculation, the strongest of our obtained cluster interactions in Ti$_{0.5}$Al$_{0.5}$N decreased with only about 2\% when the lattice parameter was expanded with 2\%. This tested expansion is substantially larger than the experimentally observed for TiN at 1692~K: 1.3\% (with respect to room temperature)~\cite{Houska1963}.
Of course, for the extreme temperatures needed to close the isostructural miscibility gap, vibrational effects are large, e.g., TiN melts at about 3500~K~\cite{Wriedt1987}. However, this region is included only for completeness and to discuss consequences of the configurational modelling scheme, not to be used for any direct comparisons with experiments.

\section{Total energy calculations and mixing enthalpies}

The total energy calculations have been done by two different electronic
structure methods. First, the electronic structure and total energies of
random c-Ti$_{1-x}$Al$_{x}$N alloys have been calculated by the exact
muffin-tin orbitals (EMTO) method in the coherent potential approximation 
(CPA) ~\cite{Vitos2001PRL,Vitos2001}. The CPA method, however, neglects local lattice  
relaxation effects and therefore, to get the relaxation energy contribution,
we combine it with supercell calculations. In the latter, the projector augmented wave (PAW)~\cite{Blochl1994} method has been used  
as implemented in the Vienna \emph{ab-initio} simulation package    
(VASP)~\cite{Kresse1996, Kresse1999}, to calculate the energies of supercells modelling the random state. These are
created by closely matching the pair correlation functions on the coordination shells where the effective cluster interactions are strongest to the values in real random alloys as described in Ref.~\cite{Alling2007}. This procedure is an extension of the strategy suggested by Zunger~\emph{et al.}~\cite{Zunger1990} in their design of so-called special quasirandom structures (SQS)~\cite{Zunger1990}, but is based on a rigorous condition for the validity of the usage of a supercell for modelling the random state~\cite{Abrikosov1997,Ruban2008REV}. Although we do not use the originally suggested small special SQS structures in this work, we use the abbreviation SQS for our random-like supercells in line with the terminology in previous publications. A comparison between SQS and Connolly-Williams cluster expansion calculations 
was recently presented by Ghosh~\emph{et~al.} for, e.g., Al-Ti intermetallics~\cite{Ghosh2008}.

Internal coordinates and volume were relaxed for each of the SQS structures while we neglected the shape relaxation, which should be absent in a 
real random alloy due to the preservation of the B1 symmetry on average.
In addition 100 ordered structures for the concentrations $x=0.25, 0.375, 0.5, 0.625$, and $0.75$, are considered 
for the cluster expansion of the lattice relaxation energies. 
The ordered structures have between 4 and 16 metal
atoms and are constructed by multiplying the unit cell in the [001], [011],
[111], and [211] directions, assuring that they all have unique combinations
of the pair correlation functions at the first two coordination shells.
Their energies have been calculated using the PAW method allowing for relaxation of all internal coordinates 
and shape, while their volumes were kept fixed at the equilibrium volume of the corresponding SQS. 
All calculations have been done using the
generalized gradient approximation (GGA)~\cite{Perdew1996} for electronic
exchange-correlation effects within the density functional framework.
The details of the calculations are
the same as in Ref.~\cite{Alling2007}.

Thus, the mixing enthalpies, at zero pressure, of the random alloys, $\Delta H_{mix}(x)$ in Eq.~(\ref{eq:H_alloy}), 
were calculated using the EMTO-CPA method for its fixed lattice part, and PAW-SQS method for the 
relaxation part, the latter decreasing the positive mixing enthalpies of the alloys with about one third~\cite{Alling2007}. Note that it was shown that the two methods gave very similar results for
fixed lattice calculations~\cite{Alling2007}.

The calculated mixing enthalpies (at zero pressure) of the SQS structures as
well as with the CPA method combined with an interpolation of the SQS lattice
relaxation energies are shown in the top panel of Fig.~\ref{fig:ordering}.
Also shown are the mixing energies, of the 100 ordered structures.

\comment{
First, we calculate the mixing energy of random alloys, $\Delta H_{mix}(c)$.
They have been calculated by using both the EMTO-CPA as well as the PAW method.
In the latter case random alloys have been model by finite size supercells,
with  distribution of alloy components, which minimizes the difference between
the supercell site occupation correlation functions and those of the random
alloy. If such model random supercells (MRS) can be used in the PAW
calculations to obtain the total mixing energy, the EMTO-CPA method provides
only it unrelaxed part, $\Delta H_{mix}^0(c)$, which is determined with atoms
in ideal lattice positions. The calculated mixing energy is shown in in the
top panel of Fig.~\ref{fig:ordering}.

{\bf Comment:} \emph{As I written in one of our papers, the term SQS is
misleading. MRS is probably misleading too, but at least there is nothing
really stupid in it like "special"}
}



\section{Effective cluster interactions}

\subsection{Effective fixed-lattice chemical interactions}

The effective fixed-lattice chemical interactions were obtained by the screened GPM
\cite{Ruban2002_a,Ruban2002_b,Ruban2004}. The starting point of this method
is the electronic structure of a random alloy at a given concentration and
external conditions. The latter is usually obtained in the CPA, which provides
the effective medium in order to calculate the energy response of the system
to a particular effective perturbation by the proper expansion
of the  one-electron energy term \cite{Ducastelle1976,Ducastelle1991}. This part is calculated within the EMTO method. In the
case of DFT formulation one then has to add an additional electrostatic
contribution following the force theorem \cite{Ruban2002_a,Ruban2002_b,Ruban2004}.
The latter contributes only to the effective pair interactions and is determined
in supercell calculations, allowing to evaluate the screening density in
the corresponding random alloy. In this work this screened electrostatic
contribution was obtained in very large random-like supercell calculations by the locally
self-consistent Green's function (LSGF) method~\cite{Abrikosov1996,Abrikosov1997}.

The screened GPM was used to calculate 1) the effective pair interactions
up to the thirtieth coordination shell, 2) all the 3-site in which the sides corresponds to
up to the fifth coordination shell, and 3) all the 4-site interactions with the
edges up to the fourth coordination shell. In addition, we also calculated
one specific, more distant, 4-site interaction, which is known to give
quite large contributions to the configurational energetics due to its specific
geometry. It is the interaction for the four site cluster along the line of the [110]
direction. The strongest of the multi-site interactions are shown in the lower panels of
Fig.~\ref{fig:Vij} 
where the interaction
index is given by the coordination shell numbers of the sides of the
corresponding cluster. In the case of the four-site interactions, the order
of indexes are of importance, so we make the following choice: the first four indexes
are the coordination shells of the sides of a closed loop through all the
four sites, and the last two are the coordination shells of remaining sides
of the cluster. 

\begin{figure}
\includegraphics[angle=-90,width=0.87\columnwidth]{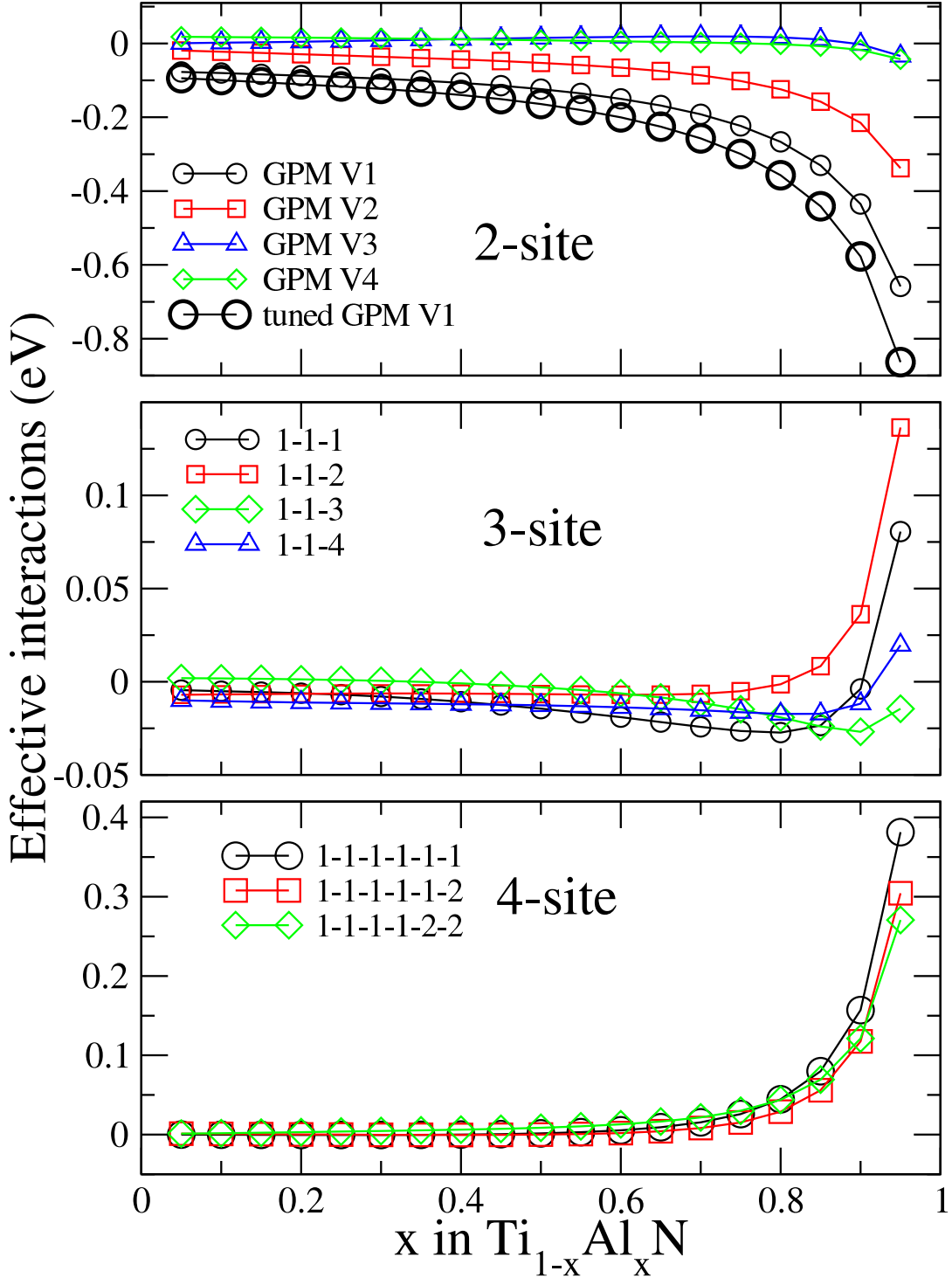}
\caption{\label{fig:Vij} (Color online) The strongest of the chemical fixed lattice effective
two-, three-, and four-site interactions as a function of Al content obtained by
the GPM method together with the tuned pair interaction at the first coordination
shell.}
\end{figure}

It is clear, that the most important effective chemical interactions of the fixed lattice are
the nearest and next-nearest neighbor pair interactions. They are negative,
favoring clustering and show a strong non-linear decrease with Al concentration.
The 3-site and 4-site interactions are all rather weak for concentrations up to
about $x \leq 0.60$. However, they increase sharply for higher Al content.
This is a signature of the gradual electronic transition where Ti 3d-states with
t$_{2g}$-symmetry, in the absence of Ti nearest (metal site) neighbors,
becomes isolated in a semiconducting AlN matrix as discussed above. It is also
the reason why it would be dangerous to expand the configurational
energy of c-TiAlN using concentration independent interaction potentials.




There are two major approximations behind the screened GPM method: the CPA and the atomic sphere approximation (ASA).
The CPA neglects the local environment effects in the electronic structure
of a random alloy. The ASA errors come in different ways, but the most 
significant one is related to the screened Coulomb interactions, which 
are defined in the specific geometry of the atomic spheres. Both these
approximations most strongly affect the interactions at the first few coordination
shells. 

In this work, we use the fixed-lattice results from the calculated mixing energies of the ordered structures to \emph{tune}
the GPM interactions. This is done by assuming that the long range pair interactions and the multi-site interactions are well described by the GPM
and then cluster expand the remaining part of the ordering energies corresponding to the short ranged
pair clusters using the concentration dependent CE method. An equivalent viewpoint is that the \emph{errors} of the GPM is cluster expanded.

In our case it turns out that it is enough to tune only one pair interaction
at the first coordination shell to obtain good agreement between the 
ordering energies from the cluster expansion and direct total-energy
calculations. It is interesting to note that the tuning results in an
almost concentration independent scaling of the GPM nearest-neighbor interaction
by factor of 1.26-1.28. This tuned-GPM interaction is shown with large circles in the top panel 
of Fig.~\ref{fig:Vij}. If we allow for simultaneous tuning also of the
next-nearest neighbor pair interactions, we obtain for this shell almost
the original GPM-value (scaling factor $\approx 1$) clearly indicating the
validity of our assumption about the short range nature of the GPM inaccuracies.
Of course, for other systems, the tuning procedure could possibly be extended
to a few more shells for the obtaining of excellent fixed-lattice interactions. 

\begin{figure}
\includegraphics[angle=-90,width=0.87\columnwidth]{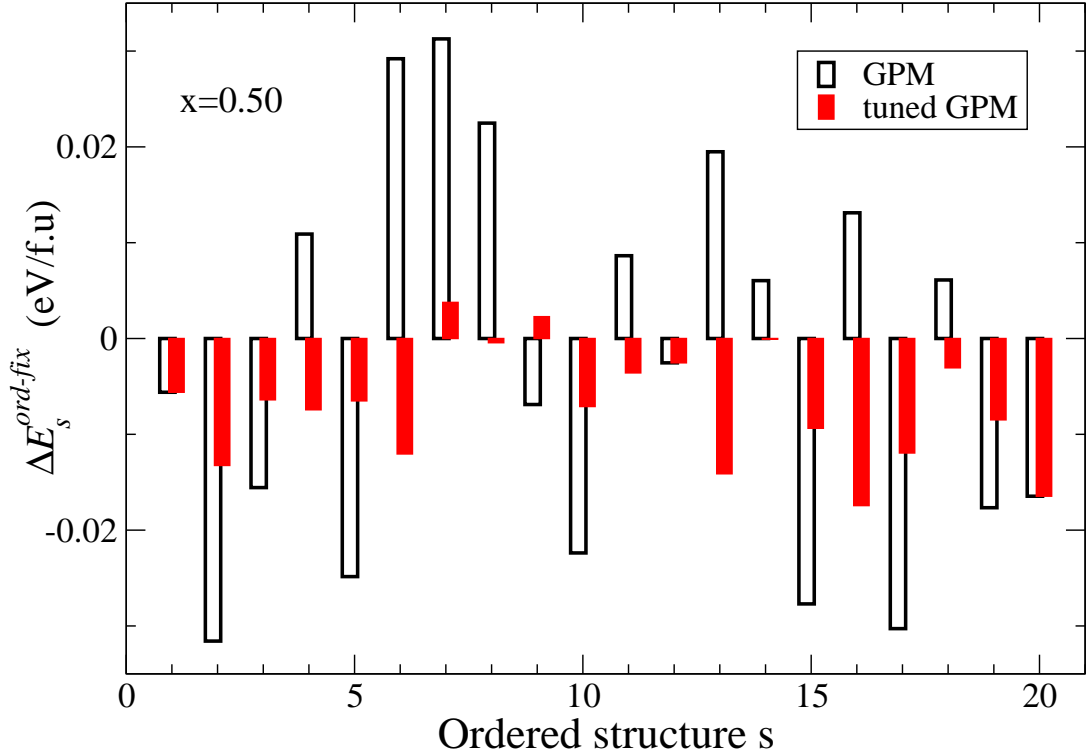}
\caption{\label{fig:delta_fix} (Color online) The error in the GPM and tuned-GPM descriptions of the fixed lattice part of the ordering energies of the 20 structures considered for Ti$_{0.5}$Al$_{0.5}$N. The average magnitude of the errors are 16.9 meV/f.u. using the GPM and 7.5 meV/f.u. using the tuned-GPM.}
\end{figure}

Figure~\ref{fig:delta_fix} shows the magnitude of the difference between the fixed-lattice ordering energies obtained with the direct PAW calculations on one hand and the 
pure screened GPM and tuned screened GPM interactions respectively on the other. This difference can be expressed as

\begin{equation}\label{eq:fix_order}
\Delta E^{s-fix} = E^{s-fix} - E^{SQS-fix} - 
\sum_f  V_f^{(n)-fix} \left( \xi_f^{(n)-s} - \xi_f^{(n)-SQS}\right)
\end{equation}
\noindent where $\xi_f^{(n)-s}=<\delta c_i \delta c_k ... \delta c_k>_f$ are
the correlation functions for figure $f$ for the alloy having structure $s$. 

 Given the large total spread in ordering energies, which is more than 250 meV/f.u for the structures with $x=0.5$ considered here, the pure GPM potentials perform reasonably well. But by using the tuning scheme, 
the accuracy becomes very good. The average absolute value of the difference is reduced from 16.9 meV/f.u. for the pure GPM, to 7.5 meV/f.u. using the tuned GPM. We note that a certain part of the remaining small error after the tuning comes from a constant shift in ordering energies. Such a small constant shift does not necessarily influence the description of energy differences between configurations treated in a statistical mechanics simulations. 


\subsection{Cluster expansion of the lattice relaxation energy}

The relaxation term of the effective interactions were obtained by
the concentration dependent Connolly-Williams CE method for the relaxation part of the energies of the ordered structures. 
The relaxation energy is defined for each structure as the energy difference between 
the situation where all atoms are sitting on ideal lattice points, and when they have been allowed to relax to
their equilibrium (eq) positions: $E^{s-rel}=E^{s-eq}-E^{s-fix}$.

 \comment{
 which is just the difference of the relaxation energies of the
ordered alloys and SQS, which are given the difference of the total energies
of the  relaxed and unrelaxed structures:
$E_{rel}^{s} = E_{tot}^{s-rel} - E_{tot}^{s-unrel}$. For each concentration
considered, $c=0.25, 0.375, 0.5, 0.625, 0.75$, we calculate the relaxation
energies of 20 different ordered structures, together with the relaxation
energy of the MRS, the latter being described in Ref.~\cite{Alling2007}.
We relax all the internal coordinates and the shape of the ordered
structures, while the volume is kept fixed at the optimal volume obtained
for the corresponding MRS. For the MRS we relax the internal coordinates
and volume, i.e. we neglected the shape relaxation, which should absent
in a real random alloy due to the preservation of the B1 symmetry on
average.
}

To obtain the $V^{(n)-rel}_\alpha(x)$ we use a least square method to minimize
the sum of the squares of the difference between the relaxation part of the ordering energies obtained
with the Ising Hamiltonian and direct DFT calculations: 
 
\begin{equation}\label{eq:order}
\Delta E^{s-rel} = E^{s-rel} - E^{SQS-rel} - 
\sum_f  V_f^{(n)-rel} \left( \xi_f^{(n)-s} - \xi_f^{(n)-SQS}\right)
\end{equation}


Of course, also our cluster expansion faces the standard obstacles of the
structure inversion method. However, we have a much easier job as compared to the
conventional cluster expansion since we have separated out the complex chemical fixed-lattice
term and only expand a part of the total ordering energies. Furthermore, judging 
from the shape of the relaxation energies of the SQS structures in Ref.~\cite{Alling2007},
 this part show no sign of peculiar concentration dependence. Instead, it was shown 
 that the local lattice relaxations in TiAlN is primarily due to relaxation of nitrogen atoms 
 positioned in between metal atoms of different chemical type, an effect mostly depending on the 
 pair correlation function on the second metal coordination shell~\cite{Alling2007}. In comparison, 
 lattice mismatch gives a quite small contribution to the relaxation energies since pure TiN ($a_{PAW}=4.255$~\AA) 
 and c-AlN ($a_{PAW}=4.07$~\AA) are rather close in lattice spacing.

For all considered concentrations, we have tested different cluster bases
with up to 12 terms, including pair interactions up to the 15:th coordination
shell, as well as short ranged (up to the first and second coordination shells)
3-, and 4-site interactions. Our conclusion is that the expansion based on the 
six pair interactions at the 1st, 2nd, 4th, 6th, 8th, and
10th shells, gives a very good description of the relaxation energies for
all concentrations. The inclusion of additional clusters, including multi-site,
in the expansion gives only minor improvements. Furthermore, the obtained
relaxation interactions depend quite weakly on the concentration, in line with
the findings in Ref.~\cite{Alling2007}. We thus perform a linear regression to
the obtained values to get the relaxation interactions for each concentration
(in steps of $\Delta x = 0.05$) to be used in the statistical mechanics simulation.

\begin{figure}
\includegraphics[angle=-90,width=0.87\columnwidth]{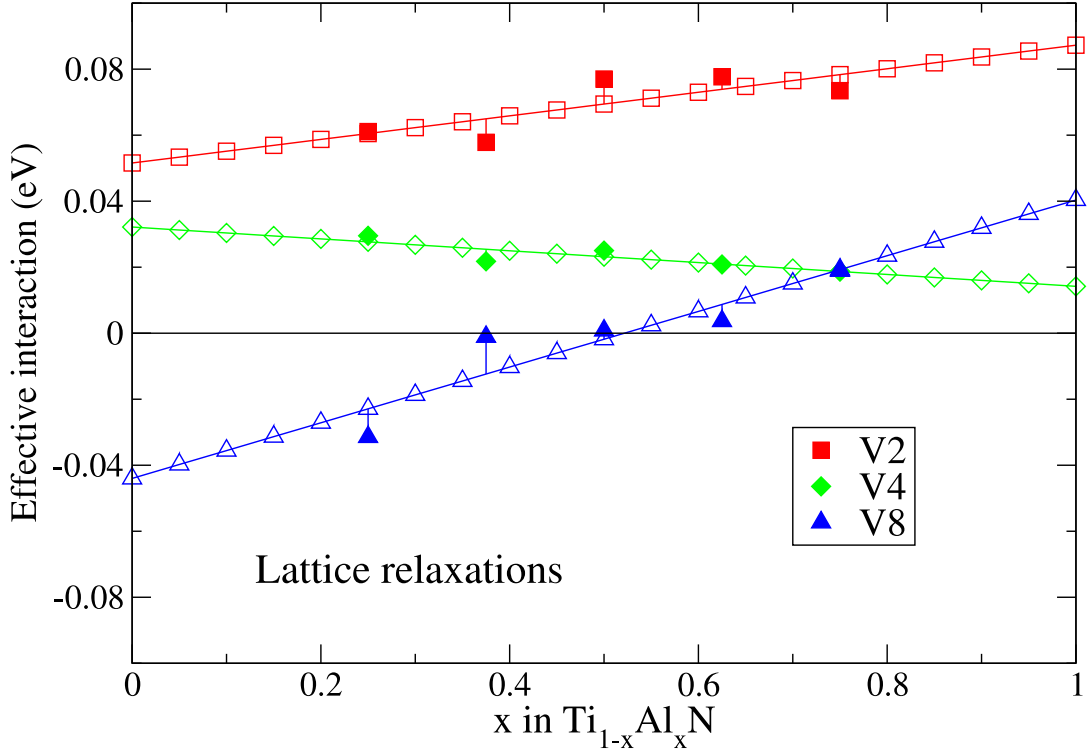}
\caption{\label{fig:relVij} (Color online) The three strongest relaxation interactions: pair interactions on the 2nd, 4th, and 8th coordination shell, as a function of Al content obtained by
the the Connolly-Williams cluster expansion method (solid symbols). The linear regression values used in the Monte Carlo simulations are shown with open symbols. Note the different scale on the y-axis as compared to the fixed lattice chemical interactions shown in Fig.~\ref{fig:Vij}}
\end{figure}

The three strongest relaxation interaction parameters, the pair interactions
on the 2nd, 4th and 8th coordination shells are shown in 
Fig.~\ref{fig:relVij} as functions of concentration together with the linear regression values 
used in the Monte Carlo simulations. One can see that the
strongest relaxation interaction, $V^{(2)-rel}_2$ is positive favoring
ordering and corresponding to the relaxation of the nitrogen positions in
between metal atoms of different kinds in line with
Ref.~\cite{Alling2007}. For all the relaxation interactions the linear regression fits well with the
obtained values. Nevertheless, we double-checked both ordering energy calculations and statistical simulations
using both the direct values and the values from the linear regression. The differences for the results were negligible.

Note that all the relaxation interactions are rather weak compared to
the first fixed lattice interaction $V^{(2)-fix}_1$, especially in
Al-rich alloys. But at lower Al content on the other hand,
the relaxation interactions become relatively more important.

The situation at $x=0.5$ is shown in Fig.~\ref{fig:V50} where the pure GPM pair
interactions are plotted with open black circles, the relaxation interactions
are shown with solid red squares and the resulting total pair interactions,
including the tuning of the fixed lattice interactions, are shown with
large, bold circles. 

\begin{figure}
\includegraphics[angle=-90,width=0.87\columnwidth]{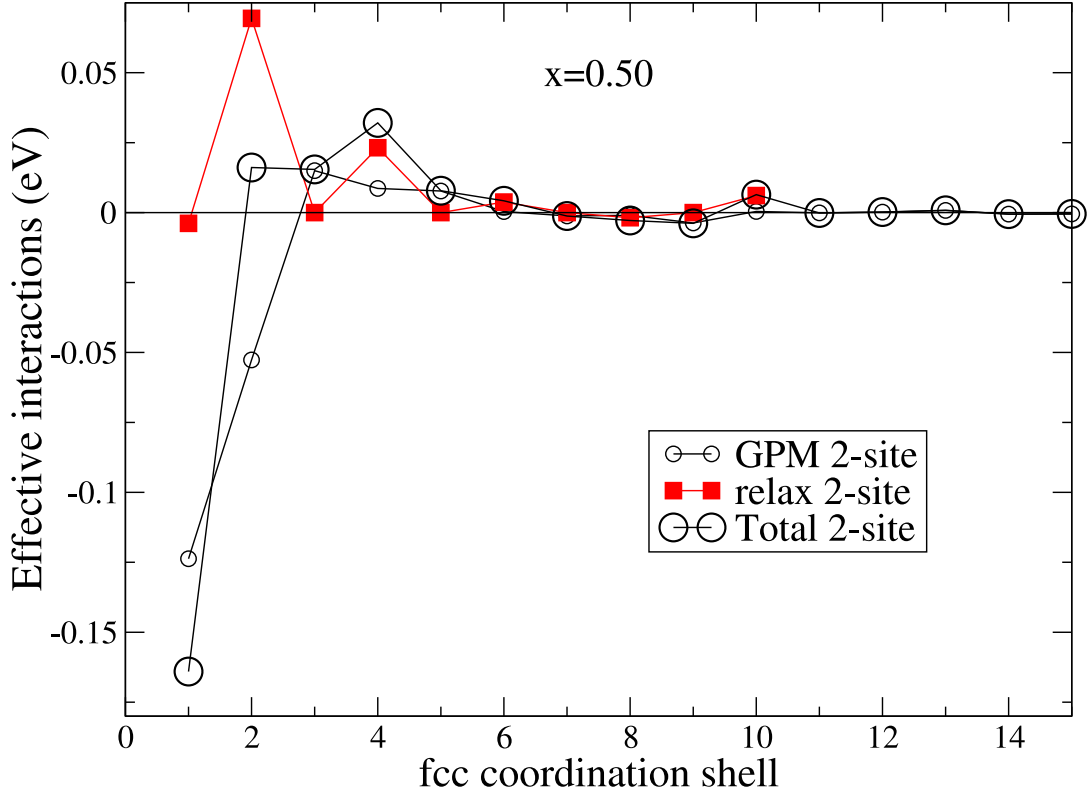}
\caption{\label{fig:V50} (Color online) The effective cluster interactions
used for the composition Ti$_{0.5}$Al$_{0.5}$N. Pure GPM interactions are shown
by small open circles, lattice relaxation interactions are shown by solid squares.
The total effective pair interactions, $V_p^{(2)}$, including also the
tuning of the fixed lattice interaction on the first coordination shell, are shown with large circles.
}
\end{figure}

\subsection{Evaluation of the effective interactions}

\begin{figure}
\includegraphics[angle=-90,width=0.87\columnwidth]{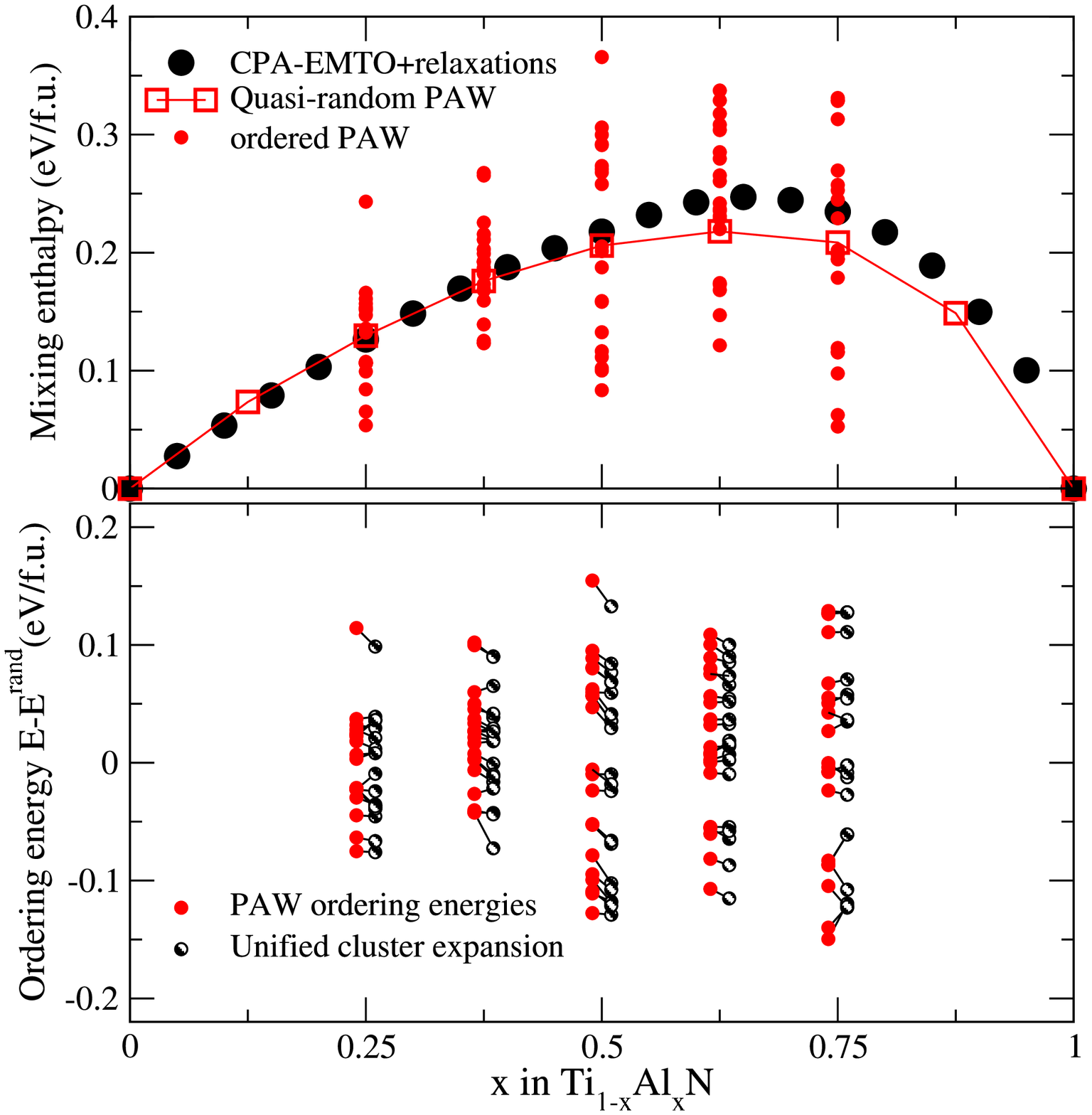}
\caption{\label{fig:ordering} (Color online) Top panel: Mixing enthalpies of Ti$_{1-x}$Al$_x$N
random alloys calculated with the CPA-EMTO method (complimented with relaxation energies) and with the 
PAW-SQS method. Also shown are the mixing energies (at the SQS volumes) for 100 ordered 
structures. 
Lower panel: The  ordering energies of the 100
different Ti$_{1-x}$Al$_x$N ordered structures.
The values of direct PAW calculations are shown with solid red circles. The unified 
cluster expansion is shown with striped black
circles. }
\end{figure}

The effective cluster interactions have been tested to produce the ordering energies
of the considered 100 structures. The results are shown in the lower panel of
Fig.~\ref{fig:ordering}. The ordering energies from
the direct calculations are based on the comparison with the SQS energies, but adjusted for the small  but
non-zero correlation functions of those supercells, $\xi_f^{(n)-SQS}$ according to 

\begin{equation}\label{eq:DFT_order}
E^{s-ord}_{direct} = E^{s}_{DFT} - \left( E^{SQS}_{DFT} - 
\sum_f  \left(V_f^{(n)-fix}+V_f^{(n)-rel} \right) \xi_f^{(n)-SQS}\right).
\end{equation}
\noindent Those results are shown with solid red circles. 

The ordering energies derived from the Ising Hamiltonian,
obtained with the unified cluster expansion method

\begin{equation}\label{eq:Ising_order}
E^{s-ord}_{Ising} = \sum_f \left( V_f^{(n)-fix}+V_f^{(n)-rel}\right) \xi_f^{(n)-s},
\end{equation}
are shown with striped circles to the right of the
results from the direct calculations. Lines connect the values obtained for the
same structure with the two different approaches. The large spread in ordering
energies, more than 0.280 eV/f.u. for structures at $x=0.5$ and almost as much for the structures with $x=0.75$, illustrate the strength
of the configurational interactions. As compared to those values the effective cluster interactions describe the ordering energies very well,
with an average magnitude of the errors of 8.8 meV/f.u. However, an increase in the errors can be seen for the structures at $x=0.75$,
 indicating the difficulties to describe the Al-richest region, as discussed above. On the other hand, this region corresponds to 
 compositions where the cubic phase can not be grown experimentally any way. Noting that we can expect a slightly larger 
 inaccuracy in the analysis of compositions with $x \geq 0.75$ we move on to the thermodynamics study.

\section{Thermodynamics of \MakeLowercase{c-}T\MakeLowercase{i}A\MakeLowercase{l}N}

\begin{figure}
\includegraphics[angle=-90,width=0.87\columnwidth]{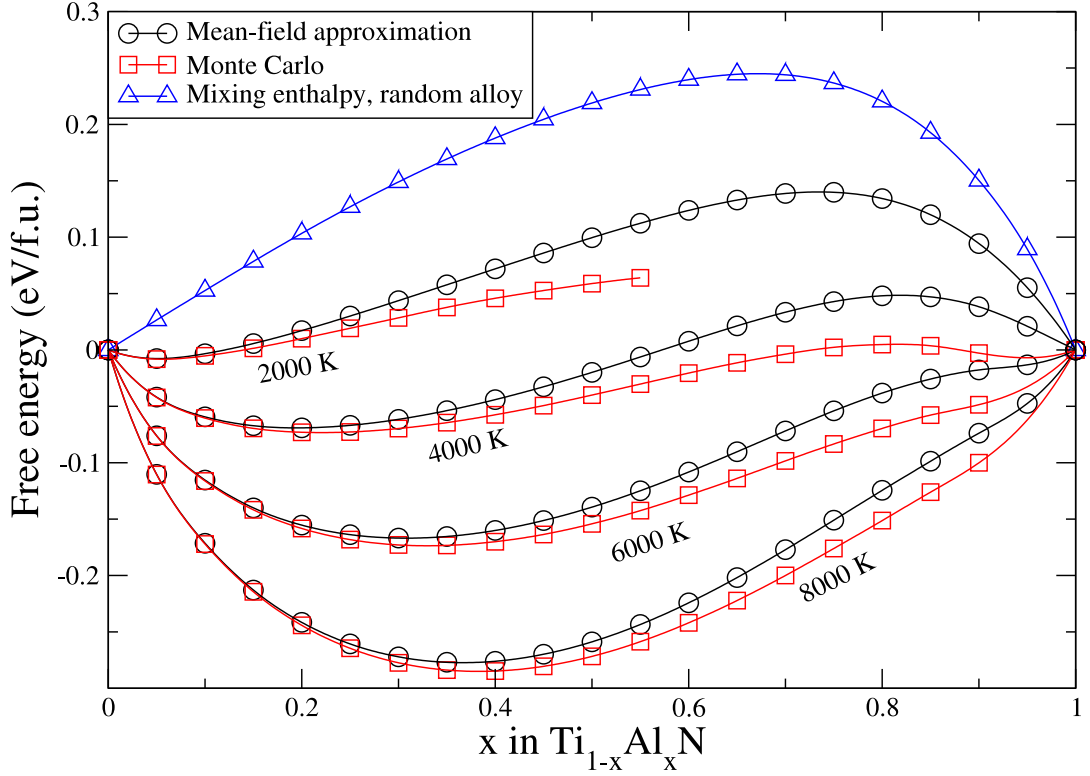}
\caption{\label{fig:free_energy} (Color online) The Gibb's free energy of mixing,
at zero pressure, for cubic Ti$_{1-x}$Al$_x$N solid solutions as calculated with
two different methods, mean-field approximation (black circles) and
Monte Carlo simulations (red squares). Values for temperatures of $T=$ 2000,
4000, 6000, and 8000~$K$ are shown together with the mixing enthalpy of the
completely random solid solution.}
\end{figure}

With the effective interaction potentials of the generalized Ising Hamiltonian
at hand we can start our analysis of the clustering thermodynamics of c-TiAlN.
The objective is to obtain the Gibb's free energy of mixing that governs the
phase stabilities. Previously, we have done so using the mean-field
approximation~\cite{Alling2007, Alling2009APL} 

\begin{equation}\label{eq:mean-field}
G(x,P,T)=H^{MF}(x,P)-TS^{MF}(x)
\end{equation}

\noindent where $H^{MF}(x,P)$ is the mixing enthalpy of the ideal random solid
solution and 
\begin{equation}\label{eq:MF-S}
S^{MF}(x)=-k_B\left(x ln x + (1-x)ln(1-x) \right)
\end{equation}
\noindent is the entropy of such a system. Since the completely random
alloy configuration has the highest entropy, mean-field
description of the configuration becomes an excellent approximation as
$T \rightarrow \infty$. At lower temperatures, short range clustering
(or ordering in other systems) will decrease the enthalpy term more than the
entropy term, so that the mean-field approximation overestimates the free
energy, and underestimates the stability of the solid solutions. 

These effects can be captured with a series of canonical Monte Carlo
calculations with the configurational Hamiltonian in Eq. (\ref{eq:H_conf}). We have
performed such calculations for fixed compositions with the step $\Delta x=0.05$
over the relevant concentration range at specific temperature. In this work we
neglect vibrational effects and consider zero pressure conditions where the
volumes used to calculate $E$ are the equilibrium volumes for each solid
solution. In this case we get

\begin{equation}\label{eq:free_energy1}
G(x,T) = E(x,T)-TS(x,T),
\end{equation} 
\noindent where
\begin{eqnarray}
E(x,T) = E^{MF}(x) + E^{MC}(x,T) \\ \label{eq:free_energy2}
S(x,T) = S^{MF}(x) + \int^T_{\infty}\frac{C_V(x,T')}{T'}dT', \label{eq:free_energy3}
\end{eqnarray} 
\noindent where $E^{MF}(x)$ is the mean-field energy, while $E^{MC}(x,T)$ and
$C_V(x,T)$ are the energy and specific heat obtained in the Monte Carlo simulation. 

We note that the Monte Carlo energy, $E^{MC}(x,T)$, is negative since it is
the deviation of the energy of the system from the energy of the ideal solid
solution due to short-range order effects at a fixed composition. Also, the
change in entropy is negative since $C_V(x,T)$ and $T$ are both positive, 
while the integration goes from higher to lower temperature values. We assume
that $S(x,T=10000~K)\approx S^{MF}(x)$ and perform the thermodynamic integration
from this temperature downwards to the temperature of interest. The resulting
free energies of mixing, calculated both within the mean-field
approximation, Eq.~\ref{eq:mean-field}, and via the more accurate treatment
from Eqs.~\ref{eq:free_energy1}-\ref{eq:free_energy3} are shown in
Fig.~\ref{fig:free_energy} for the temperature range $2000-8000~K$.

The Monte Carlo simulations are restricted by compositions $x \leq 0.9$, since at
higher Al concentrations the effective interactions become divergent and produce
quite high error. However, this does not influence more than marginally the predictions about phase
stabilities in the Ti-richer regions, which is relevant for the experimentally
achievable cubic phases. The mixing enthalpy of the completely random solid
solution, is shown for comparison.

The phase separation transition temperature can be determined for each
composition by a common tangent construction. This temperature is higher
than the temperature corresponding phase separation temperature in the 
\emph{canonical} Monte Carlo simulations. This is so since there is always
an additional energy gain for the phase separation due to the fact that
the phases, to which the phase separation undergoes in the Monte Carlo
simulations, are in a strained state due to the volume mismatch.
Thus, there is an additional contribution to the stabilization of these
phases, which is absent in the Monte Carlo simulations.

Let us note that it is difficult to perform an accurate thermodynamic
integration close to the instability temperature in the Monte Carlo
simulations. However, as discussed above, this is enough to determine both
the binodal, and for most compositions, the spinodal lines of the phase
diagram. The
Monte Carlo derived free energy curve corresponding to $T=2000~K$ is shown
only for compositions with $x\leq 0.55$ since at this temperature, the canonical clustering temperatures 
have been reached for higher Al content.

At high temperatures, e.g., $T=8000~K$, the mean-field free energy is very
close to the free energy including clustering
contribution, with only a small deviation at high Al-content where the
effective cluster interactions become quite strong. As the temperature 
decreases, the importance of the clustering becomes more and more apparent
over a larger composition range.

\begin{figure}
\includegraphics[angle=-90,width=0.87\columnwidth]{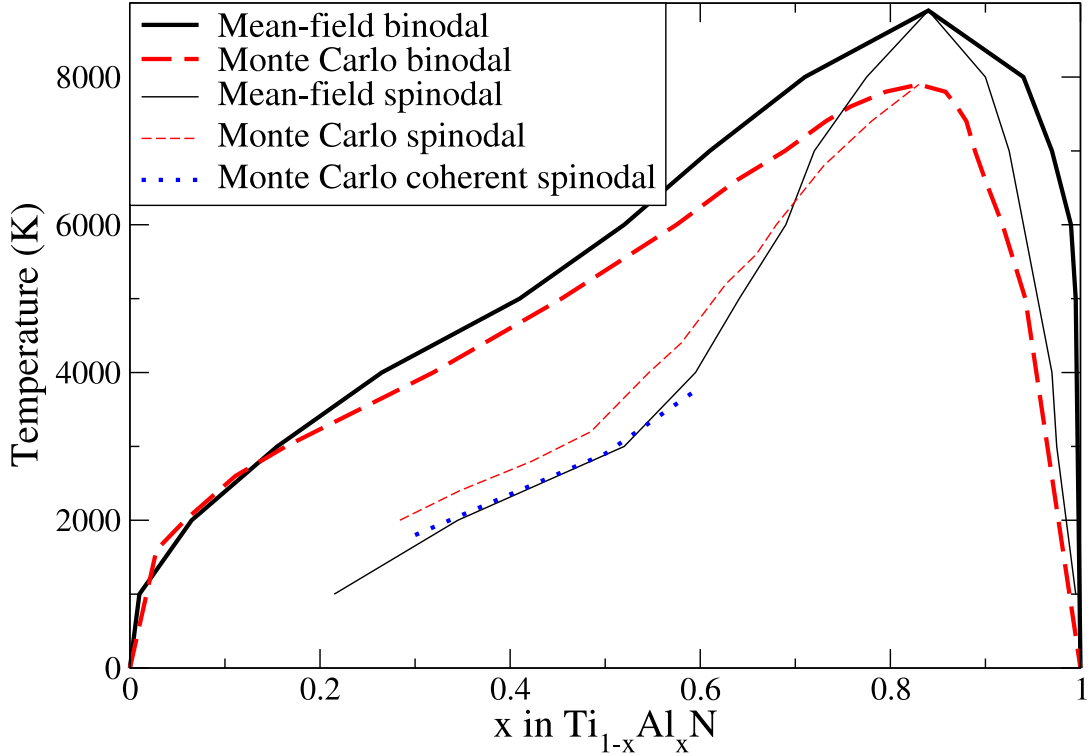}
\caption{\label{fig:phase} (Color online) The isostructural phase diagram of
cubic TiAlN as calculated with both the mean-field approximaiton,
Eq.~\ref{eq:mean-field} (black solid lines) and the Monte Carlo
approach~\ref{eq:free_energy1}-\ref{eq:free_energy3} (red dashed lines).
The binodal lines are shown with thick lines while the spinodal lines,
corresponding to the condition
$\left( \frac{\partial^2 G}{\partial x^2}\right)_T=0$, are shown by thin
lines. The simulation of the coherent spinodal is shown with a dotted line.}
\end{figure}

The phase diagram with the binodal lines derived with the common tangent
construction and the spinodal lines from the consideration of the
$\left( \dfrac{\partial^2G}{\partial x^2}\right)_T =0$ condition is seen
in Fig.~\ref{fig:phase}. The Monte Carlo derived results are actually
similar to the mean-field ones, especially in the composition region
$x\leq 0.66$, in which the single phase cubic thin films are possible to grow
experimentally~\cite{Horling2005}. In the Al-rich region, the impact of the clustering
effects are somewhat larger. The maximum temperature of the miscibility gap in
the mean-field approximation is 8900~$K$ at $x=0.84$, while the local clustering
effects reduce this temperature to 7900~$K$ at $x=0.83$. Of course, these
temperatures are of mere theoretical interest since the melting temperature
of TiN is  $\sim 3500~K$~\cite{Wriedt1987} and that the equilibrium wurtzite
structure AlN should unavoidably form at temperatures well below the closure
of the cubic miscibility gap. At temperatures of more practical interest for
the cutting applications, $T <1500~K$ the miscibility gap covers almost the
entire composition range regardless if short-range clustering effects are
considered or not.

The spinodal region, believed to be of particular importance for age-hardening,
also covers the larger part of composition space. At $T=2000~K$ compositions with
$x\geq 0.29$ are subject to spinodal decomposition according to the Monte Carlo
simulations (we neglect the Al-richest side where a minimal amount of Ti could
be present without spinodal decomposition at this temperature). In the Ti-rich
region, $x \leq 0.5$, the Monte Carlo spinodal line is actually $\sim 500~K$
above the mean field spinodal line. Since the here presented results correspond
to the so called chemical spinodal, where the free energy of each composition is
derived for its equilibrium volume, it is of interest to study also the coherent
spinodal line, the phase region where spinodal decomposition occurs even without
any possibilities for volume relaxation of the resulting phases. We estimate these
temperatures for the Al-content $x=$ 0.3, 0.4, 0.5, and 0.6 by substituting
the zero pressure mixing enthalpies with the mixing energies calculated at the
relevant fixed volume. E.g., when obtaining the coherent spinodal for $x=0.5$, all
the mixing energies entering as $\Delta H_{mix}$ in Eq.~\ref{eq:H_alloy} or
$E^{MF}(x)$ in Eq.~\ref{eq:free_energy2}, are calculated at the equilibrium
volume for c-Ti$_{0.5}$Al$_{0.5}$N. The resulting coherent spinodal is shown
with a blue dotted line in Fig.~\ref{fig:phase}. The difference from the chemical
spinodal is a decrease in spinodal temperatures by about $400-500~K$
in the here considered composition interval.

The relatively close agreement between the accurate thermodynamic treatment
and the approximate mean-field approach is quite surprising. We believe that
there is mainly two reasons for the similarities. First, in a phase separation
process the transition is governed by the competition in terms of free energy
between similar types of phases, solid solutions with different compositions.
In contrast, upon an order-disorder transition, the competition is between one
disordered and one mainly ordered phase. Thus, if the description of the
disordering is improved, such as by considering short range order/clustering
effects, the impact on the free energy difference between the competing phases
is larger in the order-disorder case, because one of the states is much more
effected than the other. On the contrary, in the phase separation case considered here,
 all the involved phases are effected in a more similar manner. 

The second reason is the presence in the TiAlN system of a small, but finite
volume mismatch. This mismatch is almost entirely governed by the composition
rather than by particular configurations at a fixed composition. Thus, this part
of the mixing enthalpies are not influenced by short range clustering, reducing
the difference between the Monte Carlo and mean-field simulations.

\section{Conclusions}
In conclusion, we have introduced the unified cluster expansion method 
and solved the difficult alloy problem of clustering thermodynamics in c-Ti$_{1-x}$Al$_x$N.
This approach illustrates how the 
two main tools of alloy theory to obtain effective cluster interactions,
the structure inversion and the generalized perturbation methods, could be fruitfully combined. 
We do so by separating the interactions into chemical
contribution obtained on a fixed lattice, for which the GPM method works well,
and the lattice relaxation term, for which the cluster expansion is applied.
Using these interactions we perform Monte Carlo simulations to determine the
short-range clustering effects on the free energy of mixing. When constructing
the isostructural phase diagram we find that cubic TiAlN is a phase separating
system over almost the whole concentration range at typical cutting tool
working temperatures $1000 \leq T \leq 1500~K$. At $T=2000~K$ the spinodal
region extends from about $x \geq 0.33$ or $x \geq 0.28$ depending if one
considers coherent decomposition conditions or not. The results show a striking,
but explainable close resemblance with those obtained with the mean-field
approximation in the composition region of experimental and industrial relevance.

\section{Acknowledgement}
The Swedish Research Council (VR), the Swedish Foundation for Strategic Research
(SSF), and the G\"oran Gustafsson Foundation for Research in Natural Sciences
and Medicine are acknowledged for financial support.
Calculations were performed using computational resources allocated by the
Swedish National Infrastructure for Computing (SNIC).


\begin{thebibliography}{31}
\expandafter\ifx\csname natexlab\endcsname\relax\def\natexlab#1{#1}\fi
\expandafter\ifx\csname bibnamefont\endcsname\relax
  \def\bibnamefont#1{#1}\fi
\expandafter\ifx\csname bibfnamefont\endcsname\relax
  \def\bibfnamefont#1{#1}\fi
\expandafter\ifx\csname citenamefont\endcsname\relax
  \def\citenamefont#1{#1}\fi
\expandafter\ifx\csname url\endcsname\relax
  \def\url#1{\texttt{#1}}\fi
\expandafter\ifx\csname urlprefix\endcsname\relax\def\urlprefix{URL }\fi
\providecommand{\bibinfo}[2]{#2}
\providecommand{\eprint}[2][]{\url{#2}}

\bibitem[{\citenamefont{Beensh-Marchwicka
  et~al.}(1981)\citenamefont{Beensh-Marchwicka, Kr{\`o}l-Stepniewska, and
  Posadowski}}]{Beensh1981}
\bibinfo{author}{\bibfnamefont{G.}~\bibnamefont{Beensh-Marchwicka}},
  \bibinfo{author}{\bibfnamefont{L.}~\bibnamefont{Kr{\`o}l-Stepniewska}},
  \bibnamefont{and}
  \bibinfo{author}{\bibfnamefont{W.}~\bibnamefont{Posadowski}},
  \bibinfo{journal}{Thin Solid Films} \textbf{\bibinfo{volume}{82}},
  \bibinfo{pages}{313} (\bibinfo{year}{1981}).

\bibitem[{\citenamefont{Knotek et~al.}(1986)\citenamefont{Knotek, B\"ohmer, and
  Leyendecker}}]{Knotek1986}
\bibinfo{author}{\bibfnamefont{O.}~\bibnamefont{Knotek}},
  \bibinfo{author}{\bibfnamefont{M.}~\bibnamefont{B\"ohmer}}, \bibnamefont{and}
  \bibinfo{author}{\bibfnamefont{T.}~\bibnamefont{Leyendecker}},
  \bibinfo{journal}{Journal of Vacuum Science and Technology A}
  \textbf{\bibinfo{volume}{4}}, \bibinfo{pages}{2695} (\bibinfo{year}{1986}).

\bibitem[{\citenamefont{M\"unz}(1986)}]{Munz1986}
\bibinfo{author}{\bibfnamefont{W.~D.} \bibnamefont{M\"unz}},
  \bibinfo{journal}{Journal of Vacuum Science and Technology A}
  \textbf{\bibinfo{volume}{4}}, \bibinfo{pages}{2717} (\bibinfo{year}{1986}).

\bibitem[{\citenamefont{Mayrhofer
  et~al.}(2006{\natexlab{a}})\citenamefont{Mayrhofer, Mitterer, Hultman, and
  Clemens}}]{Mayrhofer2006Rev}
\bibinfo{author}{\bibfnamefont{P.~H.} \bibnamefont{Mayrhofer}},
  \bibinfo{author}{\bibfnamefont{C.}~\bibnamefont{Mitterer}},
  \bibinfo{author}{\bibfnamefont{L.}~\bibnamefont{Hultman}}, \bibnamefont{and}
  \bibinfo{author}{\bibfnamefont{H.}~\bibnamefont{Clemens}},
  \bibinfo{journal}{Progress in Materials Science}
  \textbf{\bibinfo{volume}{51}}, \bibinfo{pages}{1032}
  (\bibinfo{year}{2006}{\natexlab{a}}).

\bibitem[{\citenamefont{Mayrhofer et~al.}(2003)\citenamefont{Mayrhofer,
  H\"orling, Karlsson, Sj\"ol\'en, Larsson, Mitterer, and
  Hultman}}]{Mayrhofer2003}
\bibinfo{author}{\bibfnamefont{P.~H.} \bibnamefont{Mayrhofer}},
  \bibinfo{author}{\bibfnamefont{A.}~\bibnamefont{H\"orling}},
  \bibinfo{author}{\bibfnamefont{L.}~\bibnamefont{Karlsson}},
  \bibinfo{author}{\bibfnamefont{J.}~\bibnamefont{Sj\"ol\'en}},
  \bibinfo{author}{\bibfnamefont{T.}~\bibnamefont{Larsson}},
  \bibinfo{author}{\bibfnamefont{C.}~\bibnamefont{Mitterer}}, \bibnamefont{and}
  \bibinfo{author}{\bibfnamefont{L.}~\bibnamefont{Hultman}},
  \bibinfo{journal}{Applied Physics Letters} \textbf{\bibinfo{volume}{83}},
  \bibinfo{pages}{2049} (\bibinfo{year}{2003}).

\bibitem[{\citenamefont{Santana et~al.}(2004)\citenamefont{Santana, Karimi,
  Derflinger, and Schutze}}]{Santana2004}
\bibinfo{author}{\bibfnamefont{A.~E.} \bibnamefont{Santana}},
  \bibinfo{author}{\bibfnamefont{A.}~\bibnamefont{Karimi}},
  \bibinfo{author}{\bibfnamefont{V.~H.} \bibnamefont{Derflinger}},
  \bibnamefont{and} \bibinfo{author}{\bibfnamefont{A.}~\bibnamefont{Schutze}},
  \bibinfo{journal}{Tribology Letters} \textbf{\bibinfo{volume}{17}},
  \bibinfo{pages}{689} (\bibinfo{year}{2004}).

\bibitem[{\citenamefont{Mayrhofer
  et~al.}(2006{\natexlab{b}})\citenamefont{Mayrhofer, Music, and
  Schneider}}]{Mayrhofer2006}
\bibinfo{author}{\bibfnamefont{P.~H.} \bibnamefont{Mayrhofer}},
  \bibinfo{author}{\bibfnamefont{D.}~\bibnamefont{Music}}, \bibnamefont{and}
  \bibinfo{author}{\bibfnamefont{J.~M.} \bibnamefont{Schneider}},
  \bibinfo{journal}{Applied Physics Letters} \textbf{\bibinfo{volume}{88}},
  \bibinfo{pages}{071922} (\bibinfo{year}{2006}{\natexlab{b}}).

\bibitem[{\citenamefont{Alling et~al.}(2007)\citenamefont{Alling, Ruban,
  Karimi, Peil, Simak, Hultman, and Abrikosov}}]{Alling2007}
\bibinfo{author}{\bibfnamefont{B.}~\bibnamefont{Alling}},
  \bibinfo{author}{\bibfnamefont{A.~V.} \bibnamefont{Ruban}},
  \bibinfo{author}{\bibfnamefont{A.}~\bibnamefont{Karimi}},
  \bibinfo{author}{\bibfnamefont{O.~E.} \bibnamefont{Peil}},
  \bibinfo{author}{\bibfnamefont{S.~I.} \bibnamefont{Simak}},
  \bibinfo{author}{\bibfnamefont{L.}~\bibnamefont{Hultman}}, \bibnamefont{and}
  \bibinfo{author}{\bibfnamefont{I.~A.} \bibnamefont{Abrikosov}},
  \bibinfo{journal}{Physical Review B} \textbf{\bibinfo{volume}{75}},
  \bibinfo{pages}{045123} (\bibinfo{year}{2007}).

\bibitem[{\citenamefont{Alling et~al.}(2009)\citenamefont{Alling, Od\'en,
  Hultman, and Abrikosov}}]{Alling2009APL}
\bibinfo{author}{\bibfnamefont{B.}~\bibnamefont{Alling}},
  \bibinfo{author}{\bibfnamefont{M.}~\bibnamefont{Od\'en}},
  \bibinfo{author}{\bibfnamefont{L.}~\bibnamefont{Hultman}}, \bibnamefont{and}
  \bibinfo{author}{\bibfnamefont{I.~A.} \bibnamefont{Abrikosov}},
  \bibinfo{journal}{Applied Physics Letters} \textbf{\bibinfo{volume}{95}},
  \bibinfo{pages}{181906} (\bibinfo{year}{2009}).

\bibitem[{\citenamefont{Holec et~al.}(2010)\citenamefont{Holec, Rovere,
  Mayrhofer, and Barna}}]{Holec2010}
\bibinfo{author}{\bibfnamefont{D.}~\bibnamefont{Holec}},
  \bibinfo{author}{\bibfnamefont{F.}~\bibnamefont{Rovere}},
  \bibinfo{author}{\bibfnamefont{P.~H.} \bibnamefont{Mayrhofer}},
  \bibnamefont{and} \bibinfo{author}{\bibfnamefont{P.~B.} \bibnamefont{Barna}},
  \bibinfo{journal}{Scripta Materiala} \textbf{\bibinfo{volume}{62}},
  \bibinfo{pages}{349} (\bibinfo{year}{2010}).

\bibitem[{\citenamefont{Ducastelle}(1991)}]{Ducastelle1991}
\bibinfo{author}{\bibfnamefont{F.}~\bibnamefont{Ducastelle}},
  \emph{\bibinfo{title}{Order and phase stability in Alloys}}
  (\bibinfo{publisher}{North-Holland, Amsterdam}, \bibinfo{year}{1991}).

\bibitem[{\citenamefont{Mayrhofer
  et~al.}(2006{\natexlab{c}})\citenamefont{Mayrhofer, Music, and
  Schneider}}]{Mayrhofer2006JAP}
\bibinfo{author}{\bibfnamefont{P.~H.} \bibnamefont{Mayrhofer}},
  \bibinfo{author}{\bibfnamefont{D.}~\bibnamefont{Music}}, \bibnamefont{and}
  \bibinfo{author}{\bibfnamefont{J.}~\bibnamefont{Schneider}},
  \bibinfo{journal}{Journal of Applied Physics} \textbf{\bibinfo{volume}{100}},
  \bibinfo{pages}{094906} (\bibinfo{year}{2006}{\natexlab{c}}).

\bibitem[{\citenamefont{Ducastelle and Gautier}(1976)}]{Ducastelle1976}
\bibinfo{author}{\bibfnamefont{F.}~\bibnamefont{Ducastelle}} \bibnamefont{and}
  \bibinfo{author}{\bibfnamefont{F.}~\bibnamefont{Gautier}},
  \bibinfo{journal}{Journal of Physics F: Metal Physics}
  \textbf{\bibinfo{volume}{6}}, \bibinfo{pages}{2039} (\bibinfo{year}{1976}).

\bibitem[{\citenamefont{Connolly and Williams}(1983)}]{Connolly1983}
\bibinfo{author}{\bibfnamefont{J.~W.~D.} \bibnamefont{Connolly}}
  \bibnamefont{and} \bibinfo{author}{\bibfnamefont{A.~R.}
  \bibnamefont{Williams}}, \bibinfo{journal}{Phys. Rev. B}
  \textbf{\bibinfo{volume}{27}}, \bibinfo{pages}{5169} (\bibinfo{year}{1983}).

\bibitem[{\citenamefont{Houska}(1964)}]{Houska1963}
\bibinfo{author}{\bibfnamefont{C.~R.} \bibnamefont{Houska}},
  \bibinfo{journal}{Journal of Physics and Chemistry of Solids}
  \textbf{\bibinfo{volume}{25}}, \bibinfo{pages}{359} (\bibinfo{year}{1964}).

\bibitem[{\citenamefont{Wriedt and Murray}(1987)}]{Wriedt1987}
\bibinfo{author}{\bibfnamefont{H.~A.} \bibnamefont{Wriedt}} \bibnamefont{and}
  \bibinfo{author}{\bibfnamefont{J.~L.} \bibnamefont{Murray}},
  \bibinfo{journal}{Bulletin of Alloy Phase Diagrams}
  \textbf{\bibinfo{volume}{8}}, \bibinfo{pages}{378} (\bibinfo{year}{1987}).

\bibitem[{\citenamefont{Vitos et~al.}(2001)\citenamefont{Vitos, Abrikosov, and
  Johansson}}]{Vitos2001PRL}
\bibinfo{author}{\bibfnamefont{L.}~\bibnamefont{Vitos}},
  \bibinfo{author}{\bibfnamefont{I.~A.} \bibnamefont{Abrikosov}},
  \bibnamefont{and}
  \bibinfo{author}{\bibfnamefont{B.}~\bibnamefont{Johansson}},
  \bibinfo{journal}{Physical Review Letters} \textbf{\bibinfo{volume}{87}},
  \bibinfo{pages}{156401} (\bibinfo{year}{2001}).

\bibitem[{\citenamefont{Vitos}(2001)}]{Vitos2001}
\bibinfo{author}{\bibfnamefont{L.}~\bibnamefont{Vitos}},
  \bibinfo{journal}{Physical Review B} \textbf{\bibinfo{volume}{64}},
  \bibinfo{pages}{014107} (\bibinfo{year}{2001}).

\bibitem[{\citenamefont{Bl\"ochl}(1994)}]{Blochl1994}
\bibinfo{author}{\bibfnamefont{P.~E.} \bibnamefont{Bl\"ochl}},
  \bibinfo{journal}{Physical Review B} \textbf{\bibinfo{volume}{50}},
  \bibinfo{pages}{17953} (\bibinfo{year}{1994}).

\bibitem[{\citenamefont{Kresse and Furthm\"uller}(1996)}]{Kresse1996}
\bibinfo{author}{\bibfnamefont{G.}~\bibnamefont{Kresse}} \bibnamefont{and}
  \bibinfo{author}{\bibfnamefont{J.}~\bibnamefont{Furthm\"uller}},
  \bibinfo{journal}{Phys. Rev. B} \textbf{\bibinfo{volume}{54}},
  \bibinfo{pages}{11169} (\bibinfo{year}{1996}).

\bibitem[{\citenamefont{Kresse and Joubert}(1999)}]{Kresse1999}
\bibinfo{author}{\bibfnamefont{G.}~\bibnamefont{Kresse}} \bibnamefont{and}
  \bibinfo{author}{\bibfnamefont{D.}~\bibnamefont{Joubert}},
  \bibinfo{journal}{Physical Review B} \textbf{\bibinfo{volume}{59}},
  \bibinfo{pages}{1758} (\bibinfo{year}{1999}).

\bibitem[{\citenamefont{Zunger et~al.}(1990)\citenamefont{Zunger, Wei,
  Ferreira, and Bernard}}]{Zunger1990}
\bibinfo{author}{\bibfnamefont{A.}~\bibnamefont{Zunger}},
  \bibinfo{author}{\bibfnamefont{S.~H.} \bibnamefont{Wei}},
  \bibinfo{author}{\bibfnamefont{L.~G.} \bibnamefont{Ferreira}},
  \bibnamefont{and} \bibinfo{author}{\bibfnamefont{J.~E.}
  \bibnamefont{Bernard}}, \bibinfo{journal}{Physical Review Letters}
  \textbf{\bibinfo{volume}{65}}, \bibinfo{pages}{353} (\bibinfo{year}{1990}).

\bibitem[{\citenamefont{Abrikosov et~al.}(1997)\citenamefont{Abrikosov, Simak,
  Johansson, Ruban, and Skriver}}]{Abrikosov1997}
\bibinfo{author}{\bibfnamefont{I.~A.} \bibnamefont{Abrikosov}},
  \bibinfo{author}{\bibfnamefont{S.~I.} \bibnamefont{Simak}},
  \bibinfo{author}{\bibfnamefont{B.}~\bibnamefont{Johansson}},
  \bibinfo{author}{\bibfnamefont{A.~V.} \bibnamefont{Ruban}}, \bibnamefont{and}
  \bibinfo{author}{\bibfnamefont{H.~L.} \bibnamefont{Skriver}},
  \bibinfo{journal}{Physical Review B} \textbf{\bibinfo{volume}{56}},
  \bibinfo{pages}{9319} (\bibinfo{year}{1997}).

\bibitem[{\citenamefont{Ruban and Abrikosov}(2008)}]{Ruban2008REV}
\bibinfo{author}{\bibfnamefont{A.~V.} \bibnamefont{Ruban}} \bibnamefont{and}
  \bibinfo{author}{\bibfnamefont{I.~A.} \bibnamefont{Abrikosov}},
  \bibinfo{journal}{Repports on Progress in Physics}
  \textbf{\bibinfo{volume}{71}}, \bibinfo{pages}{046501}
  (\bibinfo{year}{2008}).

\bibitem[{\citenamefont{Ghosh et~al.}(2008)\citenamefont{Ghosh, van~de Walle,
  and Asta}}]{Ghosh2008}
\bibinfo{author}{\bibfnamefont{G.}~\bibnamefont{Ghosh}},
  \bibinfo{author}{\bibfnamefont{A.}~\bibnamefont{van~de Walle}},
  \bibnamefont{and} \bibinfo{author}{\bibfnamefont{M.}~\bibnamefont{Asta}},
  \bibinfo{journal}{Acta Materialia} \textbf{\bibinfo{volume}{56}},
  \bibinfo{pages}{3202} (\bibinfo{year}{2008}).

\bibitem[{\citenamefont{Perdew et~al.}(1996)\citenamefont{Perdew, Burke, and
  Ernzerhof}}]{Perdew1996}
\bibinfo{author}{\bibfnamefont{J.~P.} \bibnamefont{Perdew}},
  \bibinfo{author}{\bibfnamefont{K.}~\bibnamefont{Burke}}, \bibnamefont{and}
  \bibinfo{author}{\bibfnamefont{M.}~\bibnamefont{Ernzerhof}},
  \bibinfo{journal}{Physical Review Letters} \textbf{\bibinfo{volume}{77}},
  \bibinfo{pages}{3865} (\bibinfo{year}{1996}).

\bibitem[{\citenamefont{Ruban and Skriver}(2002)}]{Ruban2002_a}
\bibinfo{author}{\bibfnamefont{A.~V.} \bibnamefont{Ruban}} \bibnamefont{and}
  \bibinfo{author}{\bibfnamefont{H.~L.} \bibnamefont{Skriver}},
  \bibinfo{journal}{Physical Review B} \textbf{\bibinfo{volume}{66}},
  \bibinfo{pages}{024201} (\bibinfo{year}{2002}).

\bibitem[{\citenamefont{Ruban et~al.}(2002)\citenamefont{Ruban, Simak,
  Korzhavyi, and Skriver}}]{Ruban2002_b}
\bibinfo{author}{\bibfnamefont{A.~V.} \bibnamefont{Ruban}},
  \bibinfo{author}{\bibfnamefont{S.~I.} \bibnamefont{Simak}},
  \bibinfo{author}{\bibfnamefont{P.~A.} \bibnamefont{Korzhavyi}},
  \bibnamefont{and} \bibinfo{author}{\bibfnamefont{H.~L.}
  \bibnamefont{Skriver}}, \bibinfo{journal}{Physical Review B}
  \textbf{\bibinfo{volume}{66}}, \bibinfo{pages}{024202}
  (\bibinfo{year}{2002}).

\bibitem[{\citenamefont{Ruban et~al.}(2004)\citenamefont{Ruban, Shallcross,
  Simak, and Skriver}}]{Ruban2004}
\bibinfo{author}{\bibfnamefont{A.~V.} \bibnamefont{Ruban}},
  \bibinfo{author}{\bibfnamefont{S.}~\bibnamefont{Shallcross}},
  \bibinfo{author}{\bibfnamefont{S.~I.} \bibnamefont{Simak}}, \bibnamefont{and}
  \bibinfo{author}{\bibfnamefont{H.~L.} \bibnamefont{Skriver}},
  \bibinfo{journal}{Physical Review B} \textbf{\bibinfo{volume}{70}},
  \bibinfo{pages}{125115} (\bibinfo{year}{2004}).

\bibitem[{\citenamefont{Abrikosov et~al.}(1996)\citenamefont{Abrikosov,
  Niklasson, Simak, Johansson, Ruban, and Skriver}}]{Abrikosov1996}
\bibinfo{author}{\bibfnamefont{I.~A.} \bibnamefont{Abrikosov}},
  \bibinfo{author}{\bibfnamefont{A.~M.~N.} \bibnamefont{Niklasson}},
  \bibinfo{author}{\bibfnamefont{S.~I.} \bibnamefont{Simak}},
  \bibinfo{author}{\bibfnamefont{B.}~\bibnamefont{Johansson}},
  \bibinfo{author}{\bibfnamefont{A.~V.} \bibnamefont{Ruban}}, \bibnamefont{and}
  \bibinfo{author}{\bibfnamefont{H.~L.} \bibnamefont{Skriver}},
  \bibinfo{journal}{Physical Review Letters} \textbf{\bibinfo{volume}{76}},
  \bibinfo{pages}{4203} (\bibinfo{year}{1996}).

\bibitem[{\citenamefont{H\"orling et~al.}(2005)\citenamefont{H\"orling,
  Hultman, Sj\"olen, and Karlsson}}]{Horling2005}
\bibinfo{author}{\bibfnamefont{A.}~\bibnamefont{H\"orling}},
  \bibinfo{author}{\bibfnamefont{L.}~\bibnamefont{Hultman}},
  \bibinfo{author}{\bibfnamefont{M.~O.~J.} \bibnamefont{Sj\"olen}},
  \bibnamefont{and} \bibinfo{author}{\bibfnamefont{L.}~\bibnamefont{Karlsson}},
  \bibinfo{journal}{Surface and Coatings Technology}
  \textbf{\bibinfo{volume}{191}}, \bibinfo{pages}{384} (\bibinfo{year}{2005}).

\end{thebibliography}
\end{document}